\documentclass[epj]{svjour}
\usepackage{graphics,color}
\usepackage[T1]{fontenc}
\usepackage{graphicx}
\usepackage{amsmath,amssymb,bm}
\usepackage[draft]{fixme}
\usepackage{booktabs}
\usepackage{amsmath}
\usepackage{colortbl}
\usepackage{relsize}
\usepackage{sidecap}

\usepackage{graphicx}
\usepackage{caption}
\usepackage{textcomp}
\usepackage{gensymb}

\usepackage{lipsum}

\begin{document}

\title{Complete kinematical study of the $3\alpha$ breakup of the 16.11~MeV state in $^{12}$C}
\author{K. L. Laursen\inst{1} \and H. O. U. Fynbo\inst{1}\thanks{e-mail: fynbo@phys.au.dk}  \and O. S. Kirsebom\inst{1} \and K. O. Madsb\o l\inst{1} \and K. Riisager\inst{1}
} 
\offprints{}          
\institute{Department of Physics and Astronomy, Aarhus University, DK-8000 Aarhus C, Denmark}
\date{Received: date / Revised version: date}
%
\abstract{The reaction $^{11}\text{B}+p$ has been used to populate the $(J^\pi,T) = (2^+,1)$ state at an excitation energy of 16.11\,MeV in $^{12}$C, and the breakup of the state into three $\alpha$ particles has been studied in complete kinematics. A two-step breakup model which includes interference effects is found to provide the most accurate description of the experimental data. The branching ratio to the ground state of $^8$Be is determined to be 5.1(5)\% in agreement with previous findings, but more precise by a factor of two, while the decay to the first-excited state in $^8$Be is found to be dominated by $d$-wave emission.
} 
%
\titlerunning{Complete kinematical study of the $3\alpha$ breakup of the 16.11~MeV state in $^{12}$C}
\maketitle
\section{Introduction}
\label{sec:introduction}
 

The breakup of the excited $^{12}$C nucleus into three $\alpha$ particles has been studied since the days of Lord Rutherford, motivated by a desire to understand the breakup mechanism and gain new insights into the nuclear structure~\cite{Oliphant1933}. In the 1960s and 1970s it was demonstrated that the breakup primarily proceeds in a sequential manner, {\it i.e.}, $^{12}\text{C}\rightarrow \alpha_1+{}^8\text{Be}$ followed by ${}^8\text{Be}\rightarrow \alpha_2+\alpha_3$. The sequential model was successfully applied to describe the breakup of several states in $^{12}$C~\cite{Bronson1965}, but it failed in the case of the $(J^\pi,T) = (2^+,1)$ state at an excitation energy of 16.11~MeV. Initially, this led to the suggestion that the breakup of the 16.11~MeV state proceeds directly to the 3$\alpha$ final state~\cite{Dehnhard1962}, but it was later shown that the breakup can be described within a more sophisticated sequential model, which takes into account the interference due to Bose symmetry in the $3\alpha$ final state~\cite{Chen1968,Quebert1969,Schaefer1970,Goulard1971,Hudomalj1972}.
   
The 3$\alpha$ breakup has gained renewed attention in the past decade, in part due to the advent large-area segmented silicon detectors and fast multi-channel data acquisition systems which have made it possible to collect improved experimental data. In particular, it has become possible to perform double and triple-coincidence measurements with high efficiency and high resolution (energy and angle), allowing the breakup mechanism to be studied in far greater detail than previously possible.
The renewed interest in the 3$\alpha$ breakup is also motivated by a broader interest in understanding the new multi-particle decay modes that are being discovered in exotic isotopes close to the driplines. {\it e.g.}, two-proton radioactivity. 

Modern detection techniques were first applied in the early 2000s to the breakup of the $(1^+,0)$ state at 12.71~MeV~\cite{Fynbo2003}. Using the $\beta$ decay of $^{12}$N as a means to populate the 12.71~MeV state, the breakup was measured in complete kinematics for the first time and was shown to be in quantitative agreement with a sequential model based on the $R$-matrix formalism~\cite{Bal1974}. 
More recently, the reaction $^{11}$B($^3$He$,d)$ has been used to investigate the breakups of the $(2^-,0)$ state at 11.83~MeV, the $(1^+,0)$ state at 12.71~MeV and the $(4^-,0)$ state at 13.35~MeV. Again, the same sequential model was found to provide the most accurate, though in this case not fully satisfactory, description of the breakups~\cite{Kirsebom2010}. In the same experiment the breakup of the $(0^+,0)$ state at 7.65~MeV was shown to be primarily sequential~\cite{Kirsebom2012} (see also Ref.~\cite{Itoh2014}).  

As argued in Ref.~\cite{Fynbo2009} the distinction between a sequential and a direct decay becomes ambiguous if the total decay energy is comparable to or smaller than the width of the intermediate state through which the sequential decay would proceed. In such cases it matters little which decay model one adopts. As long as Bose symmetry and spin-parity conservation are correctly incorporated into the models, the calculated $3\alpha$ momentum distributions will not differ much, making it very difficult to distinguish between sequential- and direct-decay models based on a comparison to experimental data. For some states, such as the 12.71~MeV state, the constraints imposed by Bose symmetry and spin-parity conservation are particularly strong, leaving less room for the decay mechanism to influence the $3\alpha$ final state. Indeed, fairly good descriptions of the breakup of the 12.71~MeV state have been obtained with rather different models~\cite{Fynbo2003,Kirsebom2010}, including the direct-decay model of Ref.~\cite{Korsheninnikov1990} (known as the democratic model), the three-body model of Ref.~\cite{Alvarez2007} and the aforementioned sequential model of Ref.~\cite{Bal1974}, which provides the most accurate description of the three.

Previous to this work the $3\alpha$ breakup of the $(2^+, 1)$ state at 16.11~MeV in $^{12}$C had not been studied with a modern experimental setup. (See, however, Ref.~\cite{Alcorta2012} which reports on a measurement of the branching ratio for the sequential breakup through the ground state of $^8$Be.)
The most recent studies of the breakup of the 16.11~MeV state date back to the late 1960s~\cite{Chen1968,Quebert1969} and early 1970s~\cite{Schaefer1970,Goulard1971,Hudomalj1972}.
In the present work the 16.11~MeV state is populated via the $p+{}^{11}\text{B}$ reaction, and the $3\alpha$ breakup is measured with a state-of-the-art detection system with the aim of obtaining a quantitative and accurate understanding of the breakup mechanism.  
Another aim of this work has been to measure the weak $\gamma$-decay branches of the 16.11~MeV state. Preliminary results on this aspect of the work have been published in Ref.~\cite{Fynbo2014}.
The $p+{}^{11}\text{B}$ reaction is also of interest due to its potential use as the primary source of energy in an aneutronic fusion reactor~\cite{Labaune2013}. This motivated a recent study of the $3\alpha$ breakup of the $(2^-,0)$ state at 16.6~MeV by Stave {\it et al.}~\cite{Stave2011}. 

The paper is structured in the following way: Section \ref{sec:breakup-models} describes the breakup models which will be tested against the experimental data. Section \ref{sec:exp} covers the experimental part, including a description of the setup and a discussion of the calibration procedures. Section~\ref{sec:analysis} describes the data reduction and analysis. Section \ref{sec:results} presents the results, followed by a discussion of the results in Section~\ref{sec:discussion}. Finally, Section~\ref{sec:conclusion} concludes and provides an outlook.

\section{Breakup models}
\label{sec:breakup-models}


Two conceptually different pictures of the breakup are tested in the present work: direct and sequential.

\subsection{Direct breakup}
For the direct picture we adopt the so-called democratic model of Ref.~\cite{Korsheninnikov1990}. In this model the $\alpha$-$\alpha$ interaction is assumed to play an insignificant role in the breakup, implying that the breakup proceeds without the formation of an intermediate two-body resonance. The breakup amplitude is calculated by performing an expansion in hyperspherical harmonics (eigenfunctions of the grand angular momentum operator of the the three-body system) retaining only the lowest-order term permitted by symmetries. The amplitude is further symmetrised in the coordinates of the three identical $\alpha$ particles as dictated by Bose symmetry. 

\subsection{Sequential breakup}
The sequential model takes the opposite position of the direct model. The $\alpha$-$\alpha$ interaction is assumed to play a central role in the breakup by ``locking up'' two of the $\alpha$ particles in an intermediate two-body resonance. The breakup is modelled as a sequence of two two-body breakups, {\it i.e.}, $^{12}\text{C}\rightarrow \alpha_1+{}^8\text{Be}$ followed by ${}^8\text{Be}\rightarrow \alpha_2+\alpha_3$, the only correlation between the two breakups being those due to the conservation of energy, momentum, angular momentum and parity. 
We shall refer to $\alpha_1$ as the primary $\alpha$ particle and $\alpha_2$ and $\alpha_3$ as the secondary $\alpha$ particles.
We consider breakups of the 16.11~MeV state ($J^{\pi}=2^+$) through the narrow ground state ($J^{\pi}=0^+$) and the broad first-excited state ($J^{\pi}=2^+$) in $^{8}$Be, which we shall refer to as the $^8$Be(gs) and $^8$Be(exc) channel, respectively. 
In the former case the orbital angular momenta allowed by spin-parity conservation are $l=2$ in the first decay and $l^{\prime}=0$ in the second decay; in the latter case $l = 0,2,4$ and $l^{\prime}=2$.

Implementation of the sequential model is straightforward for the $^8$Be(gs) channel, but requires special care for the $^8$Be(exc) channel due to the large width of the first-excited state in $^8$Be.
Following the approach of Refs.~\cite{Fynbo2003,Bal1974}, we employ the $R$-matrix theory~\cite{LaneThomas1958} in which resonances are parametrised in terms of level energies and reduced widths, while penetration factors account for the energy-dependent probability of quantum tunneling through the Coulomb and angular-momentum barriers.

We begin by introducing some notation,
\begin{align}
E_i &= \text{Kinetic energy of }\alpha_i\text{ in the }^{12}\text{C rest frame}\nonumber\\
E_{ij} &= \text{Relative kinetic energy of }\alpha_i\text{ and }\alpha_j \nonumber\\
(\Theta_i, \Phi_i) &= \text{Emission angles of }\alpha_i\text{ in the }^{12}\text{C rest frame}\nonumber\\
(\theta_i, \phi_i) &= \text{Emission angles of }\alpha_i\text{ in the }^{8}\text{Be rest frame}\nonumber\\
j, j^{\prime} &= \text{Total angular momentum}\nonumber\\
m, m^{\prime} &= \text{Angular momentum projection}\nonumber\\
l,l^{\prime} &= \text{Orbital angular momentum}\nonumber\\
\Gamma_l, \Gamma_{l^{\prime}}^{\prime} &= \text{Partial decay width}\nonumber\\
\gamma_l, \gamma_{l^{\prime}}^{\prime} &= \text{Reduced width}\nonumber\\
S_{l}, S_{l^{\prime}}^{\prime} &= \text{Shift function}\nonumber\\
P_{l}, P_{l^{\prime}}^{\prime} &= \text{Penetrability factor}\nonumber\\
\omega_{l}, \omega_{l^{\prime}}^{\prime} &= \text{Coulomb phase shift}\nonumber\\
\phi_{l}, \phi_{l^{\prime}}^{\prime} &= \text{Hard-sphere phase shift}\nonumber\\
E_0^{\prime} &= \text{Level energy of the }2^+\text{ resonance in }^8\text{Be} \nonumber
\end{align}
where unprimed quantities refer to the first decay, $^{12}\text{C}\rightarrow \alpha_1+{}^8\text{Be}$, and primed quantities refer to the second decay, $^{8}\text{Be}\rightarrow \alpha_2+\alpha_3$. 
Since we shall be assuming that a single orbital angular momentum dominates in the first decay, and since only a single orbital angular momentum is allowed in the second decay, we will leave out the subscripts $l$ and $l^{\prime}$ in what follows to simplify the notation. 
The partial decay widths are given by $\Gamma = 2 P(E)\gamma^2$ and $\Gamma^{\prime} = 2 P^{\prime}(E^{\prime})\gamma^{\prime 2}$, where $E = \frac{3}{2}E_1 = \frac{11}{12}E_{\text{beam}} + Q - E^{\prime}$ is the energy available in the first decay and $E^{\prime} = E_{23}$ is the energy available in the second decay, $E_{\text{beam}}$ being the kinetic energy of the proton in the laboratory frame and $Q=8.682$~MeV being the $Q$-value of the $^{11}\text{B}(p,3\alpha)$ reaction.

Disregarding the overall orientation of the breakup, knowledge of the relative kinetic energy of the secondary $\alpha$ particles, $E_{23}$, and the angle between the first and second breakup, $\theta_2$, is sufficient to fully specify the kinematics of the $3\alpha$ final state. 

\subsubsection{No symmetrisation}
Neglecting Bose symmetry and assuming that a single orbital angular momentum dominates in the first decay, the $E_{23}$ dependence of the breakup probability is given by,\footnote{The same formula appears in Ref.~\cite{Fynbo2003}, but with a wrong sign in the denominator.}
\begin{align}
\vert f \vert^2 \propto \frac{\Gamma \Gamma^{\prime}}{ \big( E_{0}^{\prime} - E_{23} - \gamma^{\prime 2}\big[ S^{\prime}(E_{23}) - S^{\prime}(E_{0}^{\prime}) \big] \big)^2 + \Gamma^{\prime 2}/4} \, .
\label{eq:rmatrix}
\end{align}
Since the reduced width in the first decay only enters as an overall multiplicative factor, not affecting the functional dependence, we arbitrarily fix it to $\gamma^2=1$~MeV. 
For the $2^+$ resonance in $^8$Be we use the $R$-matrix parameters from Ref.~\cite{Bhattacharya2006},
\begin{align}
E_0^{\prime} &= 3129\pm5\text{(stat)}\pm1\text{(sys)}\text{ keV}\;\nonumber\\
\gamma^{\prime 2} &= 1075\pm6\text{(stat)}\pm3\text{(sys)}\text{ keV}\; ,\nonumber
\end{align}
which assume a channel radius of 4.5~fm. Note that in Ref.~\cite{Bhattacharya2006} the level energy is given relative to the ground state, whereas here it is given relative to the $2\alpha$ threshold, which is 92~keV lower in energy.  
We compute the channel radii as $a=a_0(4^{\frac{1}{3}}+8^{\frac{1}{3}})$ and $a^{\prime}=a_0(4^{\frac{1}{3}}+4^{\frac{1}{3}})$ with $a_0=1.42$~fm. (This gives $a^{\prime}=4.5$~fm consistent with the channel radius adopted in Ref.~\cite{Bhattacharya2006}.)

Having assumed that a single orbital angular momentum dominates in the first decay, we can determine the $\theta_2$ dependence of the breakup probability from theory~\cite{Biedenharn1953}. (For the general case in which several orbital angular momenta contribute, the $\theta_2$ dependence cannot be uniquely determined because the relative phase shifts are not known {\it a priori}.)
Assuming that $l=2$ dominates, one obtains the following angular distribution,
\begin{equation}\label{eq:angcor}
W_{l=2}(\theta_2) = 1.12 + 0.80\sin^2 ( 2\theta_2 ) \, .
\end{equation}
Here $\theta_2$ is the angle of $\alpha_2$ relative to $\alpha_1$, measured in the $^8$Be rest frame.
As we shall see, the assumption that $l=2$ dominates is supported by the experimental data.
Finally, we note that the angular distribution for the $^8$Be(gs) channel is isotropic because the ground state has $J=0$ and hence no directional memory.

\subsubsection{Symmetrisation}

To take into account Bose symmetry, the modified expression from Ref.~\cite{Bal1974} is used for the amplitude,
\begin{align}
f_{1,23} &= {} \,  \sum\limits_{m^{\prime}}\left( l m - m^{\prime} j^{\prime} m^{\prime} \vert j m \right) 
Y_{l}^{m-m^{\prime}}\left(\Theta_1, \Phi_1 \right) 
Y_{l^{\prime}}^{m^{\prime}}\left( \theta_2, \phi_2 \right) 
\nonumber \\
&{} 
\times 
\frac{ \big[ (\Gamma/E_1^{\frac{1}{2}}) (\Gamma^{\prime}/E_{23}^{\frac{1}{2}}) \big]^{\frac{1}{2}} \, e^{i(\omega - \phi )} e^{i(\omega^{\prime} - \phi^{\prime})} }
{ E_0^{\prime} - E_{23} - \gamma^{\prime 2}\big[ S^{\prime}(E_{23}) - S^{\prime}(E_0^{\prime}) \big] - i\tfrac{1}{2}\Gamma^{\prime}} \; ,
\label{eq:bala}
\end{align}
The factors $E_1^{\frac{1}{2}}$ and $E_{23}^{\frac{1}{2}}$ have been introduced to remove the two-body phase-space factors inherent in the penetrability factors.
The breakup probability is obtained by symmetrising in the coordinates of the three $\alpha$ particles, then squaring and finally averaging over the initial spin directions,
\begin{align}\label{eq:sym}
\vert f \vert^2 = \sum\limits_{m} \vert f_{1,23} + f_{2,31} + f_{3,12} \vert^2 \, .
\end{align}
This result is then multiplied by the appropriate three-body phase-space factor.
If the symmetrisation step is neglected, Eq.~(\ref{eq:rmatrix}) is recovered.
The symmetrisation step introduces interference effects in the $3\alpha$ final state, the importance of which have been clearly demonstrated in the case of the 12.71~MeV state~\cite{Fynbo2003,Bal1974}.

\subsubsection{Coulomb repulsion}
As discussed in Ref.~\cite{Fynbo2003} it is possible to incorporate a rough correction for the Coulomb repulsion between the primary and the secondary $\alpha$ particles into the sequential model. This correction turns out to be significant for the breakup of the 12.71~MeV, where the primary $\alpha$ particle only travels a very short distance before the short-lived $^8$Be nucleus breaks up.

The correction is based on a greatly simplified picture of the breakup process, in which the $^8$Be nucleus and the primary $\alpha$ particle move apart until they reach a certain separation, $r_0$, at which point the $^8$Be nucleus breaks up.
In this picture, the primary $\alpha$ particle must first tunnel through the potential barrier of the $\alpha_1$-$^8\text{Be}$ system to $r=r_0$, after which it must tunnel through the combined potential barrier of the $\alpha_1$-$\alpha_2$ and $\alpha_1$-$\alpha_3$ systems to $r=\infty$.  
Since the tunneling probabilities combine multiplicatively, the penetrability factor in Eq.~(\ref{eq:bala}) must be replaced by,\footnote{The penetrability factor is included implicitely through $\Gamma = 2 P(E)\gamma^2$.}
\begin{equation}\label{eq:coulomb}
P_l(\tfrac{3}{2}E_1) \; \rightarrow \; \left( \frac{E_1}{E_{12}E_{13}} \right)^{\frac{1}{2}}
\, \frac{ P_l(\tfrac{3}{2}E_1) }{ \widetilde{P}_l(\tfrac{3}{2}E_1) } 
\, \widetilde{P}_{l_{12}}^{\prime}(E_{12}) 
\, \widetilde{P}_{l_{13}}^{\prime}(E_{13}) \, ,
\end{equation}
where the ``tilde'' sign indicates that the penetrability factors should be evaluated for the enlarged channel radius $\widetilde{a}=r_0$. 
For the present calculations, we adopt $\widetilde{a}=10$~fm and assume $l_{12}=l_{13}=2$ for the orbital angular momenta of the $\alpha_1$-$\alpha_2$ and $\alpha_1$-$\alpha_3$ systems. 
A na\"ive estimate of the distance travelled by the primary $\alpha$ particle may be obtained by considering the asymptotic relative speed of the $\alpha_1$-$^8\text{Be}$ system, $v\approx 0.068c$, and the mean lifetime of the first-excited state in $^8\text{Be}$, $\tau \approx 0.47\times 10^{-22}$~s, yielding $v\tau \approx 9.6$~fm. Assuming an initial separation equal to the channel radius of $a=5.1$~fm, this gives the rough estimate $r_0 \approx 15$~fm.

\subsubsection{Possible extensions of the sequential model}

Below, we outline some possible extensions of the sequential model which, however, are beyond the scope of the present study.

\paragraph{Several $\boldsymbol{l}$ values}
Eq.~(\ref{eq:bala}) is easily generalised to the case of several $l$ values by introducing a second summation running over all orbital angular momenta allowed by spin-parity conservation ($l=0,2,4$). Since, however, neither the relative magnitude nor the relative sign ($+$ or $-$) of the reduced widths, $\gamma_l$, are known, these would have to be treated as free parameters, to be constrained by fitting the experimental data.

\paragraph{Higher-lying resonances in $\boldsymbol{^8}$Be}
In addition to the two breakup channels considered here, the 16.11~MeV state could also decay via the low-energy tail of the very broad ($\Gamma \approx 3.5$~MeV) second-excited $4^+$ state in ${^8}$Be at 11.35~MeV. This channel can easily be included in the formalism, but only at the expense of introducing more free parameters. Given the good fit to the experimental data achieved with the existing model, the motivation for including the extra channel is limited.

\paragraph{Formation channel} Some degree of polarisation of the $^{12}\text{C}$ resonance formed in the $p+{}^{11}\text{B}$ reaction is to be expected. This could potentially distort the experimental Dalitz plot (cf.~Section~\ref{subsec:methodology:dalitz}) because the detection system does not cover all of 4$\pi$. An extended formalism, which takes into account the $p+{}^{11}\text{B}$ formation channel, has been developed~\cite{Goulard1970}, but introduces additional free parameters such as proton-decay widths, which would have to be constrained by fitting the experimental data.

\subsection{Summary of the breakup models}

Table \ref{tab:models} gives an overview of the four models that are being tested in the present study. M1 is a direct model based on the democratic-decay formalism of Ref.~\cite{Korsheninnikov1990} while M2--M4 are three variants of the sequential model. M2 is the unsymmetrised model based on Eq.~(\label{eq:rmatrix}) and Eq.~(\label{eq:angcor}) and assumes $l = 2$. M3 and M4 are symmetrised models based on Eq.~(\ref{eq:bala}) and Eq.~(\ref{eq:sym}), which also include the rough correction given in Eq.~(\ref{eq:coulomb}) for the Coulomb repulsion between the primary and the secondary $\alpha$ particles. M3 and M4 differ in that the former assumes $l = 0$ while the latter assumes $l = 2$.

\begin{table}[h]
\small
\centering
\caption{\label{tab:models} List of the models that are being compared to the experimental data. See the text for details.}
\begin{tabular}{cccccc} 
\toprule
Model & Seq./Demo. & Symm. & $l$ & Coulomb corr. \\
\midrule
M1 & Demo.     & yes    &  2   &  no \\
M2 & Seq.      & no     &  2   &  no \\		
M3 & Seq.      & yes    &  0   &  yes \\	
M4 & Seq.      & yes    &  2   &  yes \\
\bottomrule
\end{tabular}
\end{table}

\section{Experimental procedure}
\label{sec:exp}
The experiment was performed at the 400~keV Van de Graaff accelerator at Aarhus University. The 16.11~MeV state in $^{12}$C was populated through the $p+{}^{11}\text{B}$ reaction, using protons accelerated to energies of 167--170~keV. 
At the target position the typical beam intensity was 1\,nA, while the transversal size of the beam was approximately 2~mm $\times$ 2\,mm, as defined by a set of horizontal and vertical slits. The target consisted of natural boron on a 4~$\mu \text{g/cm}^2$ carbon backing. Several such targets were used in the experiment, with the boron thickness ranging from 10 to 15~$\mu \text{g/cm}^2$.
The reaction chamber was pumped by an oil diffusion pump. The experiment was conducted during a period of 6 months\footnote{The long measurement time was mainly motivated by the search for the weak $\gamma$-decay branches of the 16.11~MeV state.} wherein several changes were made to the setup as described below. A detailed account of the experiment is given in Ref.~\cite{laursenPhD}. 

\subsection{Detection system}
\label{subsec:exp:detector}
The detection system consisted of two double-sided silicon strip detectors (DSSSD) of the W1 type~\cite{tengblad04} with $16\times 16$ strips and an active area of 5~cm $\times$ 5~cm. The detectors used in the present experiment were both 60\,$\mu\text{m}$ thick; enough to fully stop the most energetic $\alpha$ particles from the $p+{}^{11}\text{B}$ reaction. One detector had a deadlayer of 200\,nm Si equivalent (DSSSD 1), the other 700\,nm (DSSSD 2).
\begin{figure}
\centering
\includegraphics[width=0.8\columnwidth, clip=true, trim= 50 0 60 0]{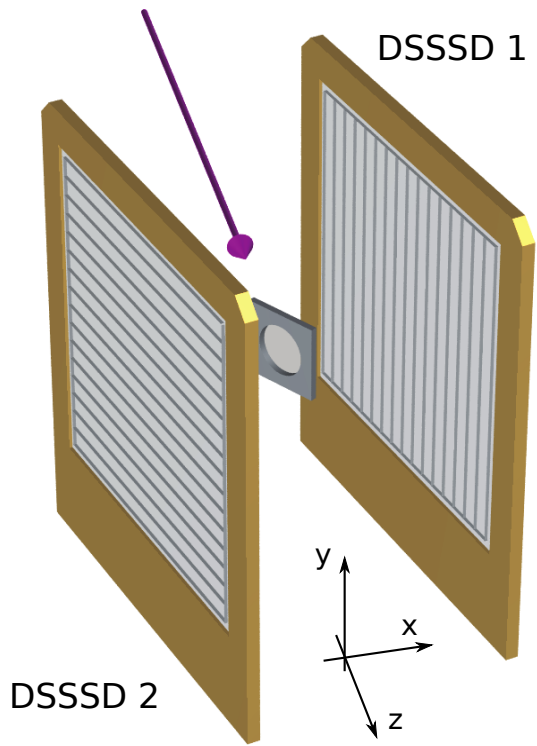}
\caption{Detector setup used in the experiment. The target is shown in the center, and the beam direction is indicated by the arrow. The active area of the detectors (light grey) measures 5~cm $\times$ 5~cm. The segmentation (dark grey) is vertical on the front and horizontal on the back.}
\label{fig:det_setup}       
\end{figure}
For the largest part of the experiment the detectors were positioned as shown in Fig.~\ref{fig:det_setup}, at a distance of 2--3~cm from the target (see Table~\ref{tab:groups} for the precise positions), providing a combined solid-angle coverage of $35\%$ of $4\pi$, with DSSSD1 covering the center-of-mass angles $60\degree$--$150\degree$ and DSSSD2 covering $35\degree$--$120\degree$. The intrinsic energy resolution of the detectors was 40\,keV (FWHM).

The setup did not allow us to discriminate between different types of particles. However, since the $3\alpha$ channel is the only open three-body channel, the $3\alpha$ events could readily be identified in the off-line data analysis as those having a multiplicity of 3. Random coincidences were identified and discarded by imposing additional cuts as discussed in detail in Section~\ref{subsec:analysis:cuts}, providing us with an efficient and highly selective method to identify the events of interest. 

The electronic signals were read out using charge-sensitive Mesytec MPR-32 preamplifiers connected to Mesytec STM16+ shaping amplifiers and analogue-to-digital-converter (ADC) modules of the CAEN 785 type. The amplification gain was stable throughout the experiment. Fast and delayed time signals, generated by the Mesytec STM16+ modules, were fed to a time-to-digital converter (TDC) of the CAEN 1190 type, providing time stamps with a resolution of about 100\,ns.

The thresholds of the data acquisition system were set as low as possible above the electronic noise level. For all electronic channels, the trigger efficiency was found to rise gently as a function of energy, increasing from 0\% to 100\% within an interval of 200 to 400\,keV, depending on the channel. Trigger thresholds, defined as the energy at which the efficiency reaches 50\%, ranged from 100 to 300\,keV for DSSSD 1 and 200 to 500\,keV for DSSSD 2. Low energy cutoffs in each ADC channel ranged from 10 to 100\,keV for DSSSD 1 and 100 to 200\,keV for DSSSD 2.
Low thresholds are essential to obtain complete kinematic information for events with low-energy $\alpha$ particles. The detection efficiency for low-energy $\alpha$ particles was further enhanced by placing the target at an angle relative to the beam axis, so that the $\alpha$ particles reaching DSSSD 2 (which has the thickest deadlayer) had to traverse the least possible amount of target material.

\subsection{Data sets}
In the course of the experiment several optimisations were made to the setup. The detectors were turned by a small angle and moved slightly closer to the target in order to achieve a better compromise between the elastic scattering rate and the solid-angle coverage. 
Small changes in the detection geometry, arising due to slight changes in the beam properties, were continuously monitored. 
The collected data has been divided into 10 data sets, each characterised by slightly different experimental conditions, as detailed in Table~\ref{tab:groups}. 
\begin{table*}[ht]
\small
\centering
\caption{\label{tab:groups} Overview of the 10 data sets. The table gives the measurement duration (col.\ 2), the beam energy (col.\ 3), the detector angles (col.\ 4 and 5), the detector positions (col.\ 6 and 7) and the target angle (col.\ 8). See Fig.~\ref{fig:det_setup} for the definition of the coordinates.}
\begin{tabular}{cccccccc} 
\toprule\toprule
Data set & Beam time [hours] & $E_\text{beam}$ [keV] & $\beta_1$ [deg] & $\beta_2$ [deg] & $(x_1, y_1, z_1)$ [mm] & $(x_2, y_2, z_2)$ [mm] & $\beta_\text{foil}$ [deg] \\
\midrule
1 & 27.5 & 167 & 110 & 290 & (20.0, -0.8, -4.5) & (-30.1, 0.0, 10.9) & 131 \\
2 & 19.3 & 167 & 110 & 290 & (20.0, -0.8, -4.9) & (-30.1, -0.5, 10.9) & 131 \\ 
3 & 17.8 & 170 & 110 & 290 & (20.0, -0.8, -9.2) & (-33.1, 0.0, 10.6) & 316 \\
4 & 15.8 & 170 & 110 & 290 & (16.9, 0.0 , -10.5) & (-32.1, 0.0, 9.0) & 306  \\
5 & 28.0 & 170 & 110 & 290 & (16.9, 0.0, -11.2) & (-32.1, 0.0, 8.8) & 306 \\
6 & 44.4 & 170 & 110 & 290 & (18.8, -0.5, -7.7) & (-24.0, -0.5, 5.8) & 306 \\
7 & 34.9 & 167 & 110 & 290 & (18.6, -0.5, -5.5) & (-23.9, -0.5, 6.1) & 126 \\
8 & 35.6 & 167 & 105 & 285 & (19.5, -0.8, -4.5) & (-24.6, -0.1, 4.7) & 126 \\
9 & 30.1 & 169 & 105 & 285 & (19.4, 0.0, -4.5) & (-23.5, -0.5, 4.90) & 126 \\	
10 & 78.9 & 169 & 105 & 285 & (19.0, 0.0, -4.9) & (-23.6, -0.5, 4.9) & 126 \\ 
\bottomrule
 \end{tabular}
\end{table*}

\subsection{Calibration}
Below, we describe the procedures adopted to calibrate the energy response and the geometry of the setup. Precise and accurate calibration is particularly important for the determination of the $3\alpha$ detection efficiency, which is highly sensitive to energy losses and thresholds effects.

\subsubsection{Geometry calibration}
The geometry is defined by specifying the position of the detectors relative to the beam spot and their orientation relative to the beam axis. 
By analyzing the hit pattern from a radioactive source placed at the target position, which emits $\alpha$ particles isotropically, the geometry can be deduced with high precision. The geometry thus obtained is, however, not entirely accurate because the source cannot be positioned exactly at the beam spot. 
This results in a distortion of the extracted kinematic curves, most easily seen in the case of the Be(gs) breakup channel which gives rise to an $\alpha$-particle group with a well-defined centre-of-mass energy of 5.8~MeV. By adjusting the geometry until the centre-of-mass energy no longer exhibits any angular dependence, we obtain a more accurate determination of the geometry, which differs by no more than 2\,mm in all three spatial directions compared to the geometry deduced from the source measurement.

\subsubsection{Energy calibration}
The six most intense $\alpha$-particle lines from the $^{228}$Th decay chain were used for the energy calibration, providing calibration points between 5.4 and 8.8~MeV. Calibrations were made at regular intervals during the experiment; only small shifts of $<0.2$\% were observed. 
SRIM range tables~\cite{SRIM} were used to correct for energy losses in the source and the detector deadlayers, taking into account the varying effective thickness due to the angle of incidence and assuming a point-like source. 
Corrections were also made for the non-ionizing energy loss \cite{Lennard1986} in the active detector volume, {\it i.e.}, the energy loss that does not contribute to the measured signal. The source thickness was determined to be 100(4)\,nm carbon equivalent by rotating the source relative to the detector while monitoring the rate of change of pulse height with angle. 
The thickness of the detector deadlayers were determined by studying the variation in pulse height across individual strips due to the changing effective thickness. 
Having corrected for the above effects, a linear fit was made to the calibration points, giving slope and offset values with statistical errors of $1\times10^{-3}$~keV/channel and 2~keV, respectively.

\subsubsection{Temporal variations}
In the course of the experiment, the energy calibration was seen to vary substantially. The variations had a recurring structure: During measurements a gradual, downward shift was observed, but when measurements were interrupted to vent the chamber, the original calibration was recovered. The largest shift observed amounted to a 90~keV decrease in the $3\alpha$ total energy.
The shift is most significant for low-energy $\alpha$ particles, suggesting that the cause is energy loss in a material which is gradually adsorbed on the target. The shift in energy calibration was found to be correlated with a gradual decline in the $3\alpha$ detection efficiency, supporting the above conclusion.
The $\alpha$-source measurements made at regular intervals did not reveal any significant changes in the calibration, which rules out adsorption on the detector surfaces. Thus, we favour the explanation that the adsorption occurs mainly on the target. The adsorped material is most likely hydrocarbons originating from the oil diffusion pump. Similar effects have been observed in other experiments employing similar pumps, see {\it e.g.}\ Ref.~\cite{kirsebom11_8B}.

For each measurement we translate the observed energy shift into an equivalent thickness of adsorped carbon. The values thus obtained range from 10 to 30~$\mu \text{g/cm}^2$. 
To keep the analysis tractable, we do not take into account the gradual nature of the absorption process when we determine the detection efficiencies. Instead we assume a constant thickness equal to half of the maximal thickness. 
The $3\alpha$ detection efficiency is, as noted above, significantly influenced by the extra energy loss in the target. This dependency it not surprising since for the 16.11~MeV state, secondary $\alpha$ particles are emitted with energies as low as 40~keV, far below the detection thresholds of our setup.

\section{Data analysis}
\label{sec:analysis}
In this section we discuss the various cuts applied to the experimental data in the off-line analysis. We also discuss how Monte Carlo simulations are used to model experimental effects and determine detection efficiencies, and finally we introduce the Dalitz-plot analysis technique.

\subsection{Data reduction}
\label{subsec:analysis:cuts}
The data reduction involves several cuts designed to remove random coincidences, {\it i.e.}, events in which one or two $\alpha$ particles from a reaction in the target are recorded in coincidence with a spurious signal due to electronic noise, an elastically scattered proton, or another $\alpha$ particle originating from a separate, but nearly simultaneous, reaction in the target. 
First, we use the TDC information to narrow the coincidence window from 2.5~$\mu$s (the width of the ADC window) to 100~ns, thereby reducing the number of random coincidences by approximately a factor of 25. 
Second, we require the energies measured on the front and back sides of the detectors to match within 150~keV, while allowing for the possibility that two particles may hit the same strip, whereby their energy is added up (summing), and the possibility that a particle may hit an interstrip region in such a way that its energy is shared between the two adjacent strips (sharing).
It may noted that summing occurs more frequently for the $^8$Be(gs) channel than the $^8$Be(exc) channel due to the small relative energy of the secondary $\alpha$ particles in the former channel. 

\begin{figure}[b]
\centering
\includegraphics[width=0.48\textwidth, clip=true, trim= 0 20 0 0]{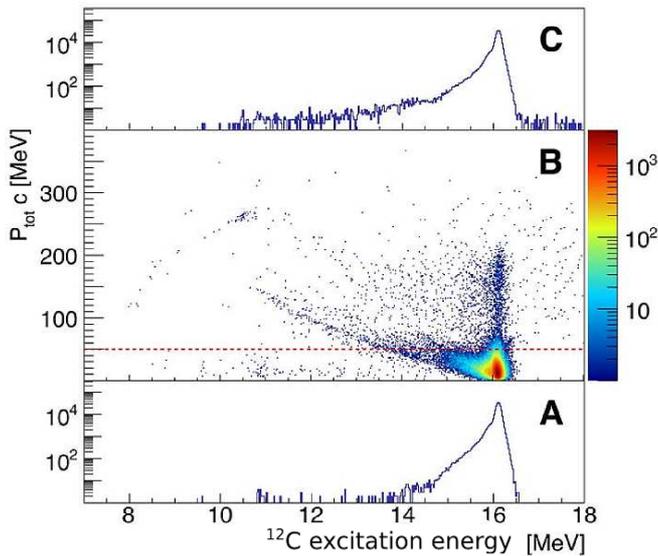}
\caption{\label{fig:pcut}(Color online) Panel B: Total momentum of the three $\alpha$ particles in the centre-of-mass frame versus the reconstructed excitation energy in $^{12}$C. (Only multiplicity-three events have been included.) Panel A: Projection with the momentum cut applied, as indicated by the dotted (red) line in panel B. Panel C: Projection onto the excitation-energy axis without the momentum cut.}
\end{figure}

We define the {\it multiplicity} of an event as the number of particles in that event which survive the above cuts. 
For those events which have a multiplicity of two, we use momentum conservation to reconstruct the momemtum of the unobserved $\alpha$ particle.
For those events which have a multiplicity of three, we can apply additional cuts to further clean the data. First, we require the total momentum to be conserved within the experimental resolution. The effect of this cut is shown in Fig.~\ref{fig:pcut}. Panel B shows the total momentum versus the excitation energy in $^{12}$C, reconstructed from the energies of the three $\alpha$ particles. Panel C shows the projection onto the excitation-energy axis without any cut imposed on the total momentum, while Panel A shows the projection obtained when we require the magnitude of the total momentum of the three $\alpha$ particles in the centre-of-mass frame to be less than 50~MeV/c, as indicated by the dotted (red) line.
In the data analysis, separate cuts are also imposed on each of the momentum components. Furthermore, we require the relative angles of the three $\alpha$ particles to add up to $360\degree$, and we require the breakup to occur in a plane. For both of these cuts a margin of $10\degree$ is allowed.

\subsection{Identification of the breakup channel}
\begin{figure}[b]
\centering
\includegraphics[width=0.48\textwidth, clip=true, trim= 0 20 20 0]{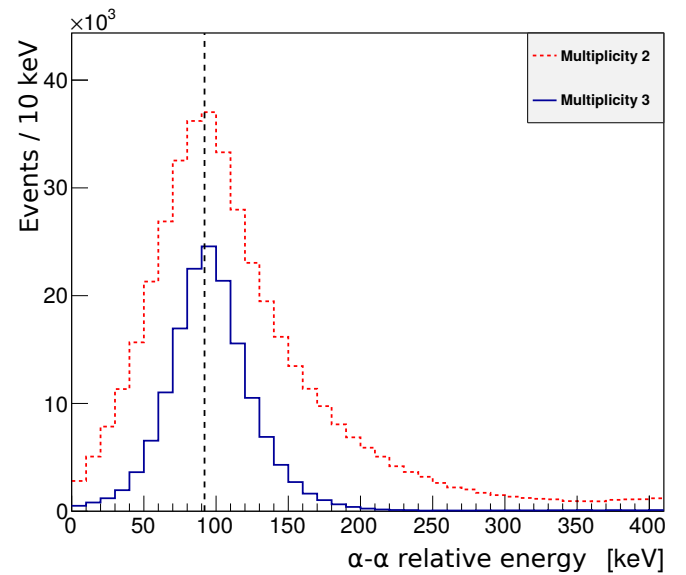}
\caption{\label{fig:pair_exc}(Color online) Relative energy of any pair of $\alpha$ particles in multiplicity-two (dotted, red) and multiplicity-three events (solid, blue) in Data Set no.\ 1. The vertical line (dashed, black) shows the $^8$Be ground-state energy relative to the $2\alpha$ threshold.}
\end{figure}

The narrow width of the $^{8}$Be ground state, combined with a high experimental resolution, makes it possible to cleanly identify the $^{8}$Be(gs) channel on an event-by-event basis by evaluating the relative energy of the three possible pairs of $\alpha$ particles,
\begin{align}
E_{ij} = \frac{(\vec{p}_i-\vec{p}_j)^2}{4M_{\alpha}} \; ,
\end{align}
where $\vec{p}_i$ and $\vec{p}_j$ are the $\alpha$-particle momenta and $M_{\alpha}$ is the $\alpha$-particle mass. 
If any pair has a relative energy consistent with the $^8$Be ground-state energy of 92~keV within the experimental resolution, we assign the event to the $^8$Be(gs) channel. In the opposite case, we assign the event to the $^8$Be(exc) channel, though this serves merely as a convenient label until we have established whether the sequential model provides an accurate description of the breakups that do not proceed through the ground state of $^8$Be.
Fig.~\ref{fig:pair_exc} shows the $\alpha$-$\alpha$ relative-energy spectrum for multiplicity-two and three events, clearly displaying the ground state peak at the expected energy.

\subsection{Experimental acceptance}
\label{subsec:analysis:spectra}

\begin{figure}[t!]
\centering
\includegraphics[width=0.48\textwidth, clip=true, trim= 0 15 10 0]{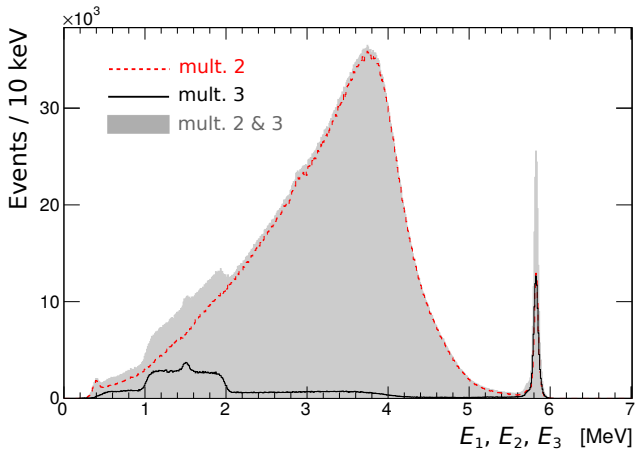}
\caption{\label{fig:e1d}(Color online) Energy spectrum measured in DSSSD~1 (Data Set no.\ 9).}
\end{figure}

The $\alpha$-particle spectrum measured in DSSSD~1 is shown by the filled histogram in Fig.~\ref{fig:e1d}, including both multiplicity-two and multiplicity-three events. The broad distribution peaking between 3 and 4~MeV and the narrow peak just below 6~MeV are the most significant structures in this spectrum. The former is the combined energy spectrum of all three $\alpha$ particles in the $^8$Be(exc) channel, while the latter is the energy spectrum of the primary $\alpha$ particle in the $^8$Be(gs) channel. 

The multiplicity-two and multiplicity-three spectra are shown separately by the solid (black) histogram and the dashed (red) histogram. The intensity of the narrow peak just below 6~MeV is similar in the two spectra, showing that for the $^8$Be(gs) channel the probability of detecting all three $\alpha$ particles is similar to that of detecting just two $\alpha$ particles. For the $^8$Be(exc) channel, on the other hand, the probability of detecting all three $\alpha$ particles is seen to be significantly reduced compared to the probability of detecting just two $\alpha$ particles. 
This difference is easily understood: For the $^8$Be(gs) channel the energy in the secondary breakup is small compared to the energy in the primary breakup, and hence the trajectories of the secondary $\alpha$ particles nearly coincidence, while for the $^8$Be(exc) channel the energies are comparable so the trajectories of the secondary $\alpha$ particles will often be very different, making the coincident detection of both secondary $\alpha$ particles unlikely. 

A Monte Carlo simulation program~\cite{alcorta2009} is used to determine the distortion of energy spectrum resulting from the limited angular coverage of the detector setup, as well as other experimental effects such as the finite beam-spot size, the energy loss in the target and the detector deadlayers, the finite granularity and intrinsic energy resolution of the detectors and the detection thresholds. 
The simulation program takes as input the $3\alpha$ final-state momentum distribution determined by the breakup models discussed in Section~\ref{sec:breakup-models}.
The ouput of the simulation is a data file with the same structure as the data collected in the experiment. This allows us to pass the simulated data through the same analysis procedure that we apply to the experimental data, thus accounting for any bias introduced by the cuts and gates applied in the analysis procedure.

\subsection{The Dalitz plot}
\label{subsec:methodology:dalitz}
Assuming an unpolarised initial state, the measurement of two $\alpha$-particle energies, $E_1$ and $E_2$, gives complete kinematic information. Thus, a two-dimensional energy plot---a so-called Dalitz plot~\cite{Dalitz}---provides a useful way to visualize the $3\alpha$ final state without loss of information. For cases such as the $3\alpha$ system, in which the masses are identical, it is advantageous to use a special version of the Dalitz plot, in which the quantities plotted on the abscissa ($\chi$) and the ordinate ($\psi$) are,
\begin{equation*}
\chi = \frac{\epsilon_1 + 2\epsilon_2 - 1}{\sqrt{3}} \; , \quad
\psi = \epsilon_1 - \frac{1}{3} \; , \nonumber
\end{equation*}
where $\epsilon_i = E_i/(E_1+E_2+E_3)$ are the $\alpha$-particle energies in the centre-of-mass frame, normalised to the total decay energy. Thus, we obtain a representation that exhibits six-fold rotational symmetry around $(X,Y)=(0,0)$ in which the kinematically allowed region is a circle with radius 1/3.
\begin{figure}[t]
\centering
\includegraphics[width=0.5\textwidth, clip=true, trim= 0 15 0 0]{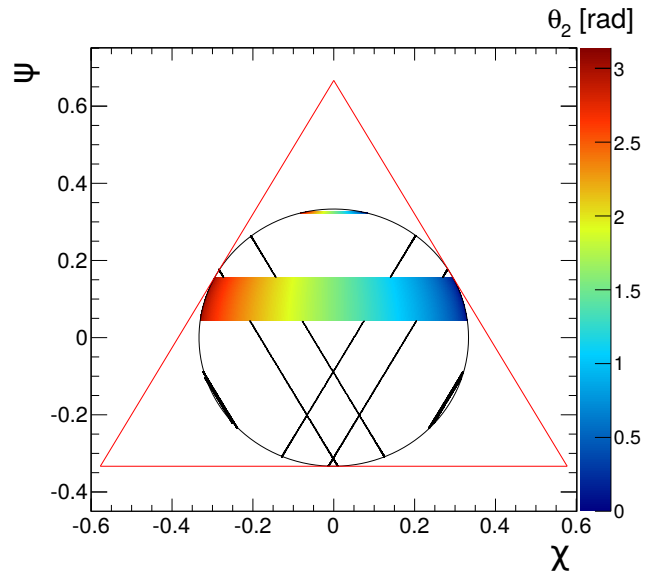}
\caption{(Color online) Schematic illustration of the Dalitz-plot distribution of a sequential breakup through the ground state (narrow bands at the rim of the circle) and the first-excited state (broad bands closer to the centre) of $^8$Be. See the text for the definition of the quantities plotted along the abscissa and the ordinate. The triangle and the circle indicate the regions allowed by energy and momentum conservation, respectively. Note that in the present figure the color scale indicates the angle between the first and the second breakup, $\theta_2$, and {\it not} the intensity.} 
\label{fig:DalitzIntro1}
\end{figure}
Since the phase-space density is constant within the kinematically allowed region, any deviation from constant density is a manifestation of symmetries in the $3\alpha$ system or dynamical correlations in the breakup process.

The Dalitz-plot distribution from a sequential breakup is shown schematically in Fig.~\ref{fig:DalitzIntro1}. The distribution is characterised by a band structure, with the $^8$Be(gs) channel producing the narrow bands near the rim of the circle, and the $^8$Be(exc) channel producing the broad bands closer to the centre. The widths of the bands reflects the widths of the intermediate two-body resonances. 
In the full $R$-matrix description, the intensity distribution {\it across} the bands reflects the profile of the intermediate two-body resonance, modified by the penetration factors in the entrance and exit channels, while the intensity distribution {\it along} the bands reflects the angular-correlation function, as seen from the color scale in Fig.~\ref{fig:DalitzIntro1}. Finally, we note that interference effects due to Bose symmetry are expected where the bands overlap which, as seen in Fig \ref{fig:DalitzIntro1}, only occurs for the $^8$Be(exc) channel.

\section{Results}
\label{sec:results}

\subsection{Dalitz distribution of the $^8$Be(exc) channel}
\label{subsec:resultsdalitz}

\begin{figure*}[ht!]
\centering
\includegraphics[width=1.0\textwidth, clip=true, trim= 0 10 0 0]{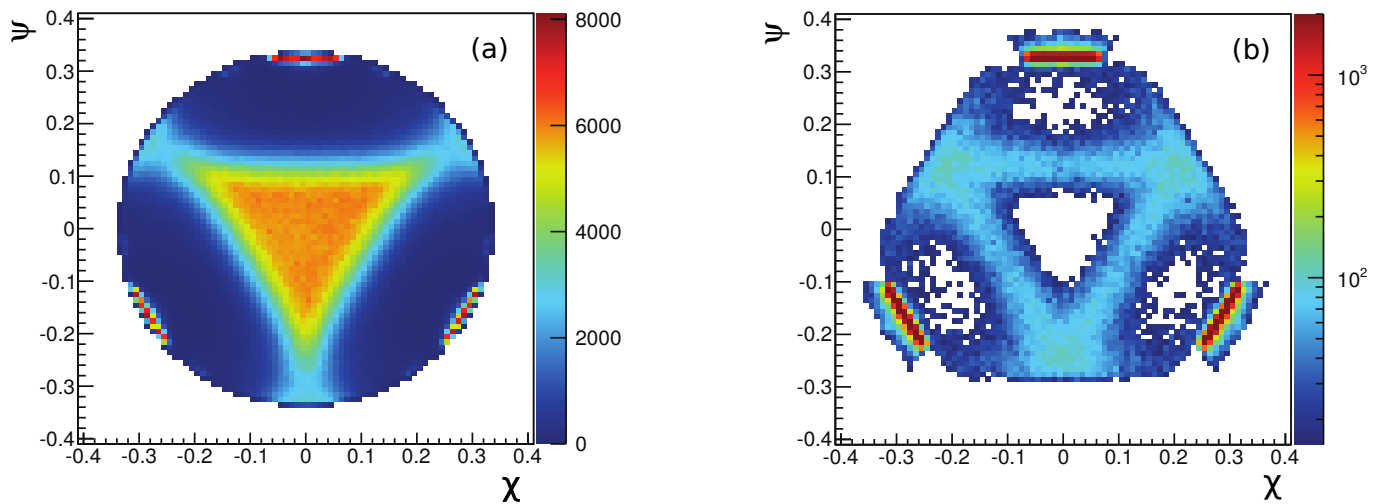}
\caption{\label{fig:dalitz_meas}(Color online) Dalitz distribution measured in the present experiment (Data Set no.\ 9), separated into multiplicity-two (a) and multiplicity-three (b) events. The color scale shows the event density.}
\end{figure*}

The Dalitz distribution of the 16.11~MeV state measured in the present experiment is shown in Fig.~\ref{fig:dalitz_meas}, separated into multiplicity-two events (a) and multiplicity-three events (b). As discussed in Section~\ref{subsec:analysis:spectra}, the difference between the two distributions is entirely an effect of the experimental acceptance. 
The lack of events near the centre of the multiplicity-three distribution reflects the fact that for the $^8$Be(exc) channel the the probability of detecting all three $\alpha$ particles is significantly reduced compared to the probability of detecting just two $\alpha$ particles.
In contrast, no such suppression is observed for the $^8$Be(gs) channel (the three narrow bands near the circumference of the Dalitz plot), reflecting the fact that for the $^8$Be(gs) channel the probability of detecting all three $\alpha$ particles is similar to that of detecting just two $\alpha$ particles.

\begin{figure}[ht!]
\centering
\includegraphics[width=0.45\textwidth]{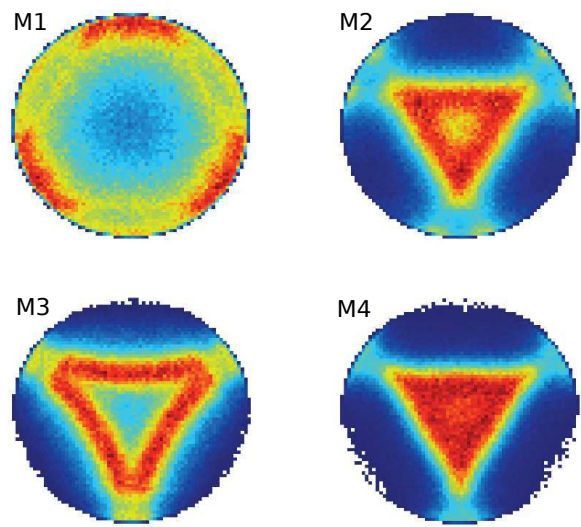}
\caption{\label{fig:dalitz_4plots}(Color online) Dalitz distributions generated from simulations of multiplicity-two events based on the breakup models discussed in Section~\ref{sec:breakup-models}.}
\end{figure}

In the following we focus on the $^8$Be(exc) channel, which is much richer in physics than the $^8$Be(gs) channel due to the large width and non-zero spin of the first-excited state in $^8$Be.
Multiplicity-two Dalitz distributions generated from simulations of the $^8$Be(exc) channel are shown in Fig.~\ref{fig:dalitz_4plots}.
The different breakup models (M1--M4) give noticeably different distributions. By comparing to the measured distribution, shown in Fig.~\ref{fig:dalitz_meas}~(a), we conclude that M4 provides the most accurate description of the breakup. 
The democratic model (M1) fails altogether at reproducing the triangular shape of the measured distribution, whereas the sequential models (M2--M4) all reproduce it in various degrees. %
Among the sequential models, M3, which assumes an $s$-wave ($l=0$) primary $\alpha$ particle, is the least consistent with the measured distribution, while M2 and M4 both come quite close, showing that the breakup is dominated by a $d$-wave ($l=2$) primary $\alpha$ particle. M4, which includes symmetrisation, is seen to fill out the inner region in better agreement with the measured distribution than M2, which does not include symmetrisation. Thus, the effect of the symmetrisation is to cause constructive interference at the centre of the triangle and destructive interference on the outside, resulting in sharper edges and a more uniform  intensity distribution within the triangle.

\begin{figure*}[ht!]
\centering
\includegraphics[width=1.0\textwidth]{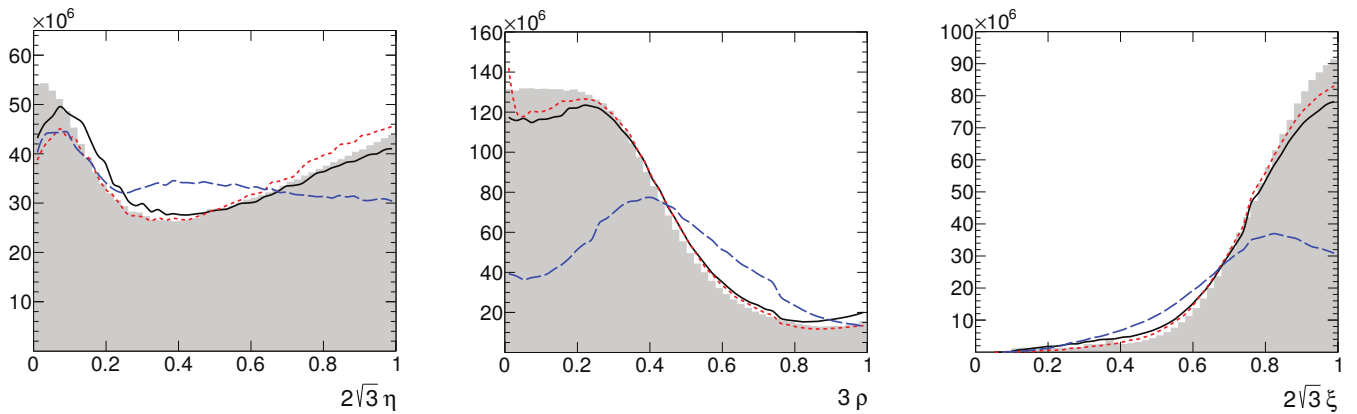}
\caption{\label{fig:dalitz_proj_meas}(Color online) Projections of the multiplicity-two Dalitz distribution. The filled histograms (grey) show the measured data. The curves show the simulated data based on model M2 (blue long dashed), M3 (black solid) and M4 (red dashed).}
\end{figure*}

To facilitate a quantitative comparison of the simulated and measured data, we consider three different projections of the Dalitz plot, designed to highlight different aspects of the two-dimentional distribution. The projected coordinates $\rho$, $\xi$ and $\eta$ are given by~\cite{Kirsebom2013},
\begin{align}
(3\rho)^2 &= (3\epsilon_i - 1)^2 + 3(\epsilon_i + 2\epsilon_j - 1)^2 \nonumber \\
2\sqrt{3}\xi &= 1-2(\epsilon_i - \epsilon_j) \label{eq:proj_coor} \\
2\sqrt{3}\eta &= 3-2(\epsilon_i + 2\epsilon_j) \nonumber \; , 
\end{align}
where we have re-ordered the $\alpha$-particle energies such that $\epsilon_i < \epsilon_j < \epsilon_k$.
The projections thus obtained are shown in Fig.~\ref{fig:dalitz_proj_meas}. In accordance with our previous conclusion, M3 and M4, which both assume $l=2$, give the most accurate description of the experimental data. A close comparison of M3 and M4, which only differ by the extra barrier-penetrability factors included in M4, reveals that M3 gives a slightly better description of the $\eta$ projection, while M4 gives the best description of the $\rho$ and $\xi$ projections.

\subsection{Branching ratio of the $^8$Be(gs) channel}

In order to extract a precise and accurate value for the branching ratio of the $^8$Be(gs) channel, precise and accurate knowledge of the coincidence detection efficiency for both the $^8$Be(gs) and the $^8$Be(exc) channel is necessary.
We use our Monte Carlo simulation program to determine the detection efficiencies for both channels. For the $^8$Be(exc) channel we adopt the breakup model M4, since it was found to give the best fit to the measured Dalitz distribution.
Detection efficiencies are determined separately for each of the 10 data sets listed in Table~\ref{tab:groups}.
The multiplicity-three detection efficiencies thus obtained range from 11 to 15\% and from 0.2 to 0.8\% for the $^8$Be(gs) and $^8$Be(exc) channels, respectively. The corresponding efficiencies for multiplicity two range from 23 to 30\% and from 17 to 29\%.
Correcting for the efficiencies, we determine the branching ratio of the $^8$Be(gs) channel to be $5.4(1.1)\%$, using multiplicity-two data, $5.1(5)\%$, using multiplicity-three data. The two values are mutually consistent, and furthermore they are consistent with the most recent literature value of $5.8(9)\%$~\cite{Alcorta2012}, with our multiplicity-three value being more precise by almost a factor 2. 
Note that these values do not include the contribution due to the ghost of the $^8$Be ground state~\cite{Barker1962}, which was found to be about 20\% in Ref.~\cite{Alcorta2012}.

\section{Discussion}
\label{sec:discussion}

In Section~\ref{subsec:resultsdalitz} we showed that a sequential breakup model, which includes Bose symmetry and a rough correction for final-state Coulomb repulsion, gives a reasonable fit to the experimental data. 
In contrast, the democratic, direct-decay model was found to give a poor fit to the experimental data.
The simple picture of a stepwise breakup thus appears to provide a fairly accurate description of the breakup of the 16.11~MeV state in $^{12}\text{C}$.
Our ability to clearly discriminate between the two breakup mechanisms hings on the fact that the total decay energy ($E=8.8$~MeV) is significantly larger than the width of the first-excited state in $^8$Be ($\Gamma^{\prime}=1.5$~MeV). For lower-lying states in $^{12}\text{C}$, the distinction is much less clear~\cite{Fynbo2009}.

Our observation that $d$-wave emission dominates in the first decay, $^{12}\text{C}\rightarrow \alpha_1+{}^8\text{Be}$, is in accordance with the observations of Refs.~\cite{Quebert1969,Schaefer1970,Goulard1971,Hudomalj1972}. It is remarkable that $d$-wave emission is so strongly favoured over $s$-wave emission, given that both decays occur above the barrier (the mean decay energy is $5.8$~MeV, while the barrier heights for the $s$- and $d$-wave channels are 2.2~MeV and 4.0~MeV, respectively) and hence neither is inhibited by barrier penetration. A similar  observation has been made for the $2^-$ state at $16.57$~MeV where $f$-wave ($l = 3$) is favoured over $p$-wave ($l = 1$) \cite{Quebert1969,Stave2011}.

The small (5\%) branching ratio of the $^8$Be(gs) channel is another surprising feature of the breakup of the 16.11~MeV state. Considering only barrier penetrability factors, one would expect a branching ratio of 60\% for the $^8$Be(gs) channel with the $^8$Be(exc) channel accounting for the remaining 40\%.

Given that barrier penetration cannot explain the $d$-wave dominance nor the slowness of the ground-state transition, we conclude that these anomalies are caused by the structure of the 16.11~MeV state, in particular its overlap with the first-excited state in $^8$Be.
This naturally leads to the question of how the 16.11~MeV state can decay to three $\alpha$ particles in the first place, considering that it belongs to an isopin triplet ($T=1$). The $\alpha$ decay must occur through admixtures of one or several nearby $(J^{\pi},T)=(2^+,0)$ states. The bound $(2^+,0)$ state at 4.44~MeV and the giant quadropole resonance around 26 MeV have previously been suggested as candidates~\cite{Lind1977}, but in recent years evidence has been found for several, hitherto unknown, low-lying $(2^+,0)$ states in $^{12}$C~\cite{Hyldegaard2009,Freer2012,Zimmerman2013,Zimmerman2013b}, providing additional candidates. It would be interesting to study the isospin mixing between these states and the 16.11~MeV state with modern microscopic cluster models.

\section{Conclusions and outlook}
\label{sec:conclusion}

The present high-statistics measurement of the $3\alpha$ breakup of the $(J^\pi, T)=(2^+,1)$ state at 16.11~MeV in $^{12}$C provides the most accurate understanding of the decay mechanism to date.
A sequential model, which assumes a stepwise decay through the two lowest-lying resonances in $^8$Be, is found to provide a rather accurate description of the breakup. 
Quantitative agreement with the experimental data is only obtained if Bose symmetry is included in the model. The agreement is further improved, though only slightly so, by including a rough correction for final-state Coulomb repulsion. In the end very good agreement is obtained though small systematic deviations remain.

The branching ratio to the ground state of $^8$Be is determined to be $5.1(5)\%$ in good agreement with previous findings, but more precise by a factor of two, and the decay to the first-excited state in $^8$Be is found to be dominated by $d$-wave emission, also in agreement with previous findings. 
It is conjectured that these non-intuitive properties of the breakup are a consequence of the structure of the 16.11~MeV state, or more precisely, the structure of one or several $(2^+,0)$ states that are mixed into the 16.11~MeV state, enabling the decay into three $\alpha$ particles.
The experimental and analytical methods used to investigate the breakup of the 16.11~MeV state here, can be applied directly to other resonances in the $p+{}^{11}\text{B}$ reaction, {\it e.g.}, the resonance associated with the $(2^-,0)$ state at 16.6~MeV, the breakup of which has recently been studied with a somewhat simpler detector set-up and analysis method~\cite{Stave2011}.

\section*{Acknowledgements} 
This work has been supported by the European Research Council under ERC starting grant LOBENA, No. 307447. OSK acknowledges support from the Villum Foundation.\\


\end{document}